# Strong coupling superconductivity at 8.4 K in an antiperovskite phosphide SrPt$_3$P

T. Takayama[1], K. Kuwano[1], D. Hirai[1], Y. Katsura[2], A. Yamamoto[2] and H. Takagi[1,2,3]

[1]Department of Advanced Materials, University of Tokyo, and JST-TRIP, Kashiwa, Chiba 275-8651, Japan

[2]RIKEN Advanced Science Institute, Wako, Saitama 351-0198, Japan

[3]Department of Physics, University of Tokyo, Hongo, Tokyo 113-0033, Japan

## Abstract

We report the discovery of a family of ternary platinum phosphides APt$_3$P (A = Ca, Sr and La), which crystallize in an antiperovskite-based structure closely related to that of the heavy fermion superconductor CePt$_3$Si. All three phosphides showed superconductivity at low temperatures and the highest critical temperature $T_c$ = 8.4 K was observed for SrPt$_3$P. The analysis of specific heat $C(T)$ for SrPt$_3$P shows clear evidence for very strong coupling $s$-wave superconductivity with a large ratio between superconducting gap $\Delta_0$ and $T_c$, $2\Delta_0/k_BT_c \sim 5$, and the presence of low-energy phonons. The presence of multiple Fermi surface pockets was inferred from the nonlinear magnetic field dependence of Hall resistivity, which we argue might play a role in realizing the strong coupling of charge carriers with the low lying phonons.

The discovery of high-$T_c$ superconductivity in iron pnictides [1] provided a fresh fuel to the exploration of new superconductors. There seems to be more than one route to realize superconductivity at a high temperature. The close proximity to a magnetically ordered state has been widely believed to be one of the most promising route, which is the case for superconductivity in cuprates, Fe pnictides, organics and heavy fermions. Electron-phonon mediated superconductivity might have been once abandoned but, with the discovery of MgB$_2$ with $T_c$ = 39 K[2], revived as another promising route. In the case of MgB$_2$, it was discussed that charge carries in strongly covalent B $2p_\sigma$ bands enhances a coupling to high energy optical phonons, resulting in relatively high-$T_c$[3]. Similar scenario might be drawn for superconductivity in doped diamonds[4]. Low-lying phonon such as rattling mode is known to enhance electron-phonon coupling $\lambda = N(0)<I^2>/M<\omega^2>$ and hence $T_c$, which is discussed to be the case for KOs$_2$O$_6$[5], and Chevrel phase[6]. In these compounds, the ratio between superconducting gap and transition temperature $2\Delta_0/k_BT_c$, a good measure of such strong coupling, is anomalously large, close to 5. Because of the lack of superconductors falling into this category, we yet do not have clear guiding principle to design such a situation.

In the course of search for high-$T_c$ superconductors with the above in mind, we discovered a family of new platinum-based phosphides APt$_3$P (A = Sr, Ca and La), which shows superconductivity at $T_c$ = 8.4 K, 6.6 K and 1.5 K, respectively. In this letter, we report the discovery with emphasis on the unique nature of the superconductivity. Those phosphides have a distorted antiperovskite-based structure, which are related to a heavy fermion superconductor CePt$_3$Si but are non-polar unlike CePt$_3$Si. We will show evidences for strong coupling $s$-wave superconductivity in SrPt$_3$P with the highest $T_c$ = 8.4 K and will discuss the presence of low lying phonons and the multiple Fermi surface pockets as a possible key ingredient for the very strong-coupling with $2\Delta_0/k_BT_c \sim 5$.

The samples used in this study were polycrystalline powders prepared by a conventional solid state reaction. Powders of elemental platinum, red phosphorus and alkaline earth (lanthanum) were mixed in an argon-filled glove box, and sealed in a quartz tube filled with argon gas. The tube was initially heated up to 400 ˚C and held at the temperature for 12 h in order to avoid rapid volatilization of phosphorus, subsequently calcined at 900 ˚C for 72 h. The sintered pellet was reground and further annealed at 900 ˚C within an argon-filled quartz tubes for several days and finally quenched into ice water. The structural analysis was conducted by x-ray diffraction (XRD) using Cu-$K\alpha$ radiation (Rigaku, RINT-UltimaIII). Magnetic, electric and thermodynamic properties were measured using commercial apparatus (Quantum Design, MPMS and PPMS).

A signature of superconductivity around 8 K was captured at the very early stage of the study in Sr-Pt-P ternary powder samples with a starting composition of Sr : Pt : P = 1 : 2 : 2. The x-ray diffraction (XRD) pattern indicated that these superconducting samples were a mixture of different phases including $PtP_2$, small signature of $ThCr_2Si_2$-type phase and other unknown phases. By repeating the synthesis with changing the starting compositions and carefully examining the phases showing up in the XRD pattern and the diamagnetic response in the superconducting state, the superconducting phase was identified to be $SrPt_3P$ with a tetragonal unit cell with $a$ = 5.809 Å, $c$ = 5.383 Å. By synthesizing from 1 : 3 : 1 mixture, indeed, we obtained the almost single phase of $SrPt_3P$ with tiny amount of unknown phases, whose XRD profile is shown in Fig. 1. Since none of known structures to date could satisfactory account for the XRD profile, we used an *ab*-initio structural analysis algorithm, "charge-flipping"[7], to construct a structural model with the aid of superflip software[8]. Using the unit cell parameters and peak intensities of the XRD pattern, the charge density map was computed[9]. Based on that, we constructed an initial structural model and refined it by Rietveld technique using RIETAN-2000 software[10]. The result of refinement showed a reasonably good convergence (reliability factors; $R_{wp}$ = 14.00%, $S$ = 1.51) and provided structural parameters listed in Table 1.

The structure of $SrPt_3P$ is pictured in the inset of Fig. 1, which consists of alternative stacking of layer of distorted anti-perovskite $Pt_6P$ octahedral units and layer of Sr. This structure is closely related to the structure of a heavy fermion superconductor without inversion symmetry, $CePt_3Si$[11]. Unlike $CePt_3Si$, the polarity of asymmetric distortion of octahedra in $SrPt_3P$ alternates within the *ab*-planes, forming an "antipolar" pattern. An inversion center therefore exists in $SrPt_3P$. Due to the antipolar arrangement, a $\sqrt{2}a_p \times \sqrt{2}a_p$ supercell is formed in the *ab* plane, where $a_p$ denotes the primitive antipervoskite cell length. To the best of our knowledge, this is the first example in which a phosphorus atom is encapsulated in an antiperovskite octahedron.

Resistivity measurements show that $SrPt_3P$ is a metal and is superconducting with $T_c$ = 8.4 K. As shown in Figs. 2 (a) and 4(b), a zero resistance and a large diamagnetic signal were clearly observed below $T_c$. The diamagnetic shielding and Meissner fractions were 97% and 35%, respectively, hallmarking bulk superconductivity of $SrPt_3P$. Further evidence for bulk superconductivity was obtained from the large specific heat jump at $T_c$ shown in Fig. 2(b). The magnetization curve in the superconducting state showed a typical behavior of type-II superconductors. The upper critical field $\mu_0H_{c2}(T)$ in the inset of Fig. 2(a), evaluated from the midpoints of resistive transitions, showed an almost linear temperature dependence down to the lowest temperature. $\mu_0H_{c2}(0)$ was estimated to be around 5.7 T, which yields a Ginzburg-Landau coherence length $\xi_{GL}(0)$ ~76 Å. Note that the linear behavior of $\mu_0 H_{c2}(T)$ down to low temperatures deviates appreciably from those expected from the Werthamer-Helfand-Hohenberg formula[12]. We suspect that a strong spin-orbit coupling effect inherent to heavy 5$d$ Pt might be involved to give rise to this behavior.

The normal state specific heat $C_N(T)$ was estimated by applying a magnetic field $\mu_0 H$ = 9 T, as shown in the inset of Fig. 2(b). The $C_N(T)/T$ vs $T^2$ plot below 10 K showed a pronounced non-linear behavior, very likely implying the presence of low lying phonons of meV scale. We therefore fitted $C_N(T)$ with $C_N(T) = \gamma T + \beta T^3 + \delta T^5$, which yielded $\gamma$ = 12.7 mJ/mol·K$^2$, $\beta$ = 1.29 mJ/mol·K$^4$ and $\delta$ = 7.98×10$^{-3}$ mJ/mol·K$^6$ [13]. The Debye temperature $\Theta_D = (12\pi^4 NR/5\beta)^{1/3}$ was ~190 K, which is comparable to that of non-centrosymmetric antiperovskite superconductor LaPt$_3$Si($T_c$= 0.6 K, $\Theta_D$ ~ 170 K). In LaPt$_3$Si, however, the signature of low lying phonons as observed in SrPt$_3$P is lacking[14].

Based on the normal state specific heat $C_N$ and $\gamma$ obtained above, the electronic specific heat under zero field $C_{el}(T)$ was estimated as shown in Fig. 2 (b) where $\Delta C$ represents the difference between specific heats under 0 and 9 T ($\Delta C/T = C(T)/T - C_N(T)/T = C_{el}(T)/T - \gamma$). The behavior of $C_{el}(T)$ below $T_c$ clearly demonstrates a strong coupling superconductivity with a finite energy gap, very likely $s$-wave symmetry in SrPt$_3$P. The electronic specific heat $C_{BCS}(T)$ expected for BCS weak coupling limit is also shown as dotted line for comparison. It is clear that the decrease of $C_{el}(T)$ at low temperatures is much more rapid than $C_{BCS}(T)$, indicating the presence of a finite superconducting gap appreciably larger than the weak coupling limit value 1.76$k_B T_c$, and therefore a strong coupling $s$-wave superconductivity. In accord with this, the specific heat jump at $T_c$, $\Delta C/\gamma T_c$, apparently exceeds 2 and is much larger than the BCS weak coupling limit value of 1.42. By employing the so-called $\alpha$-model[15] with a dimensionless parameter $\alpha = \Delta_0/k_B T_c$ representing the coupling strength, the experimental data can be indeed well fitted with $\alpha = \Delta_0/k_B T_c$ ~ 2.55 ($\Delta_0$ = 1.85 meV and $T_c$ = 8.40 K). The obtained 2$\Delta_0/k_B T_c$ ~ 5 largely exceeds the weak coupling BCS value of 3.52. The strong coupling $s$-wave superconductivity with 2$\Delta_0/k_B T_c$ ~ 5 is rather rare, only a few examples known to date include Pb-Bi alloy, A15 compounds[16], pyrochlore osmates[5], and Chevrel phases[6].

The normal state magnetic susceptibility, shown in Fig. 3(b), consists of a temperature independent contribution, which can be ascribed to the summation of those from Pauli paramagnetism, van Vleck paramagnetism and core diamagnetism, and Curie-like contribution, very likely from magnetic impurities. By subtracting the Curie-like contribution, we estimate the temperature independent magnetic susceptibility as $\chi_0$ ~ 0.5×10$^{-4}$ emu/mol. The core diamagnetism is estimated to be ~ $-1.4\times 10^{-4}$ emu/mol by using those reported for $Sr^{2+}$, $Pt^{2+}$ and $P^{3+}$. This provides an upper limit of the Pauli paramagnetic susceptibility $\chi_P = 1.9\times 10^{-4}$ emu/mol, which, combined with $\gamma = 12.7$ mJ/mol·$K^2$, yields an upper limit of the Wilson ratio $R_W = \pi^2 k_B^2 \chi_P /3\mu_B^2\gamma \sim 1$, the free electron value. The van Vleck paramagnetism is large for Pt compound because of the strong spin-orbit coupling and is often of the order of $10^{-4}$ emu/mol[17] but is hard to estimate quantitatively. Considering the van Vleck term, the Wilson ratio is likely much smaller than 1. This is in marked contrast with strongly correlated systems, where $R_W$ is close to 2. Considering the absence of signature of profound electron correlations, it may be reasonable to conclude that the pairing in $SrPt_3P$ is mediated by phonons, likely the low lying phonons demonstrated by the specific heat. $R_W$ smaller than 1 is not surprising in the presence of strong electron-phonon coupling. In a strong coupling superconductor such as Pb, $R_W$ is usually much less than 1 (For Pb, $R_W$ is estimated to be ~ 0.1) because the specific heat coefficient $\gamma$ is the subject of enhancement due to electron-phonon coupling, as $\gamma = (1+\lambda)\gamma_b$ where $\gamma_b$ represents the bare density of states, whereas the Pauli susceptibility $\chi_P$ is not.

The evidence of strong coupling of charge carriers with low lying phonons may be captured in the normal state resistivity $\rho(T)$ shown in Fig. 3(a). $\rho(T)$ shows almost $T$-linear resistivity down to very low temperatures below 20 K, which implies the presence of low energy (~ meV) modes responsible for scatterings. These low energy modes can be reasonably ascribed to low lying phonons observed in the specific heat. With increasing temperature, we see very clearly that the resistivity increase tends to saturate above 100 K. This signals that the electron mean free path is approaching to the order of atomic lattice spacing, known as resistivity saturation. The resistivity saturation was observed in strong coupling superconductors such as A15 compounds [18] and was interpreted as an evidence for the strong coupling of charge carriers with phonons.

The presence of multiple Fermi surfaces in $SrPt_3P$ is suggested from the magnetotransport measurements demonstrated in Fig. 3(c). The Hall resistivity is strongly temperature and field dependent, indicative of the presence of two type of carries with distinct mobilities[19]. The strongly non-quadratic behavior of magnetoresistance at high fields and at low temperatures, where the non-linear behavior of Hall resistivity is significant, is consistent with such multiple carrier situation. The crossover of the field dependence of Hall resistivity at 10 K, from with large positive slope (~ $1\times 10^{-3}$ $cm^3$/C) in the low field limit to negative slope in the high fields, implies the coexistence of small hole pocket(s) containing of the order of $10^{21}$ carriers of high mobility and lower mobility carriers in electron pocket(s). We suspect that strong spin-orbit coupling inherent to $5d$ Pt might play some role in creating Fermi surface pockets by splitting the degenerate $d$ bands.

CaPt$_3$P and LaPt$_3$P were also successfully synthesized in almost single phases including small traces of Pt metal and PtP$_2$, and confirmed to crystallize in the same crystal structure as SrPt$_3$P by XRD, with the lattice constants $a$ = 5.6673(2) Å, $c$ = 5.4665(3) Å for CaPt$_3$P and $a$ = 5.7626(3) Å, $c$ = 5.4650(4) Å for LaPt$_3$P, respectively. Both CaPt$_3$P and LaPt$_3$P were found to show superconductivity reproducibly at around 6.6 K and 1.5 K, respectively, as demonstrated both by the resistive and the diamagnetic transitions in Fig. 4. The specific heat jump is seen at $T_c$, though it is broadened appreciably as compared with that of SrPt$_3$P, very likely reflecting poorer homogeneity of sample. The electronic specific heat $\gamma$ were estimated to be 17.4 and 6.7 mJ/mol·K$^2$ for CaPt$_3$P and LaPt$_3$P, respectively. It is interesting to infer here that $\gamma$ for CaPt$_3$P is substantially larger than that of SrPt$_3$P, which cannot be accounted for even if the difference in the quality of sample is taken into account. The fact that $T_c$ is not simply scaled by the electronic density of states might suggest that the coupling strength depends strongly on the details of structure. Another interesting point we should raise here is the contrasted behavior of resistivity in LaPt$_3$P. As clearly seen in Fig. 3(a), in the high temperature side, we do not see clear signature of resistivity saturation up to room temperature and, in the low temperature side, the resistivity crosses over from $T$-linear to a higher power-law dependence at much higher temperature than SrPt$_3$P[20]. The behavior is much more like a conventional metal with weak electron-phonon coupling, which is in agreement with much lower $T_c$ for LaPt$_3$P than SrPt$_3$P. Note that, since LaPt$_3$P accommodates one more valence electron than SrPt$_3$P, the topology and the orbital character of Fermi surface should be drastically different from SrPt$_3$P.

The experimental results on SrPt$_3$P presented so far point to the fact that the charge carriers, accommodated in the multiple Fermi surface pockets, couple very strongly with the low lying phonons and that strong coupling superconductivity therefore is realized at relatively high temperature close to 10 K. Though speculative at this stage, it might be interesting to point out here that the presence of multiple Fermi surface pockets might be enhancing the electron-phonon coupling through increasing the likelihood for the Fermi surface nesting and the resultant softening of phonons. The contrasting superconducting properties between SrPt$_3$P and LaPt$_3$P, with different Fermi surface topologies, are consistent with this scenario, which is worthy for being examined both theoretically and experimentally.

In summary, a new family of platinum phosphides superconductors was discovered. The structure can be viewed as an anti-polar analogue of that of CePt$_3$Si, a "non-centrosymmetric" heavy fermion superconductor, which might give rise to a unique opportunity to capture the influence of lack of inversion symmetry if we are able to synthesize both polar and anti-polar structures consisting of the electronically equivalent elements. The evidence for unexpectedly strong electron-phonon coupling was found in the highest-$T_c$ compound of the family, SrPt$_3$P. We propose that the low lying phonons, arguably produced by and coupled strongly with the multiple Fermi surface pockets, might be key ingredients in realizing strong coupling superconductivity at relatively high temperature.

We thank M. Nohara, S. Shamoto, D. Hashizume and A. Bangura for stimulating discussions. This work was partly supported by Grant-in-Aid for Scientific Research (S) (Grant No. 19104008), Grant-in-Aid for Scientific Research on Priority Areas (Grant No. 19052008).

**Figures**

Fig. 1 (color online) X-ray diffraction pattern of SrPt$_3$P at room temperature registered with a Cu-$K\alpha$ radiation. The black cross represents the experimental data, and the red solid line shows the calculated pattern. The green bars indicate the expected peak positions based on the structural model. Inset: Crystal structure of APt$_3$P. Gray, red and yellow spheres represent platinum, phosphorus and A (=Sr, Ca, La) atoms, respectively.

Fig. 2 (color online) Superconducting properties of SrPt$_3$P displaying (a) resistive transition and (b) specific heat divided by temperature associated with superconductivity. The solid line represents the fitting curve based on the $\alpha$-model[15], and the dotted one shows that of the weak-coupling BCS model $C_{BCS}$. Insets: (a) Temperature dependence of the upper critical field $\mu_0 H_{c2}(T)$ evaluated from magnetoresistive curves. (b) Temperature dependence of normal state specific heat divided by temperature. The solid line indicates the fitting line, and the dotted one shows a linear fit as a guide for eyes.

Fig. 3 (color online) Normal state properties of SrPt$_3$P displaying temperature dependences of (a) resistivity, (b) magnetic susceptibility and (c) Hall coefficient. In (a), resistivity of LaPt$_3$P is also presented. Insets: Isotherm (a) magnetoresistances and (c) Hall resistivities. The red, yellow, green and blue plots represent the data registered at 280, 100, 50 and 10 K, respectively.

Fig. 4 (color online) Superconducting transitions of APt$_3$P (A = Ca, Sr and La) in (a) normalized resistivities, (b) Meissner effects and (c) heat capacities.

Table 1 Structural parameters of SrPt$_3$P refined by a Rietveld analysis. The space group is $P4/nmm$ (no. 129) and $Z = 2$, and the lattice constants are $a = 5.8094(1)$ Å and $c = 5.3833(2)$ Å. Throughout the refinement, the isotropic atomic displacement parameters are fixed at 0.1 Å$^2$.

| atom | site | g | x | y | z |
|---|---|---|---|---|---|
| Pt(1) | 4e | 1.0 | 1/4 | 1/4 | 1/2 |
| Pt(2) | 2c | 1.0 | 0 | 1/2 | 0.1409(3) |
| Sr(1) | 2a | 1.0 | 0 | 0 | 0 |
| P(1) | 2c | 1.0 | 0 | 1/2 | 0.7226(16) |

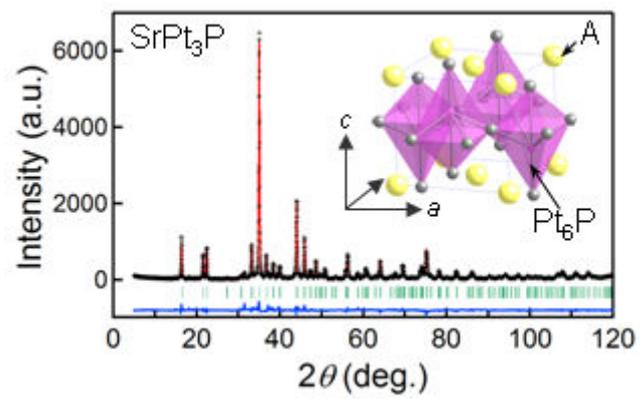

**Fig. 1**

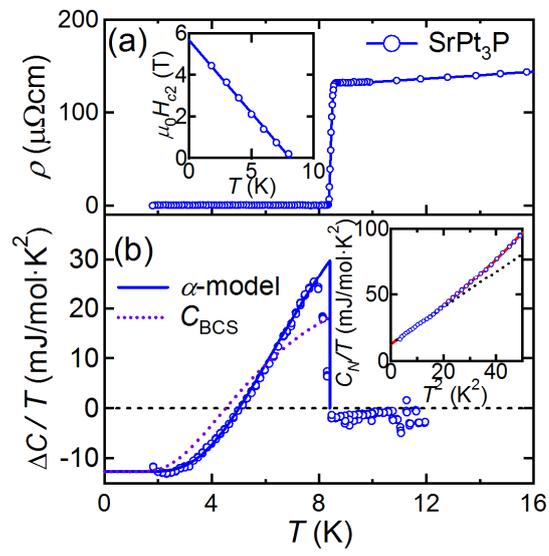

Fig. 2

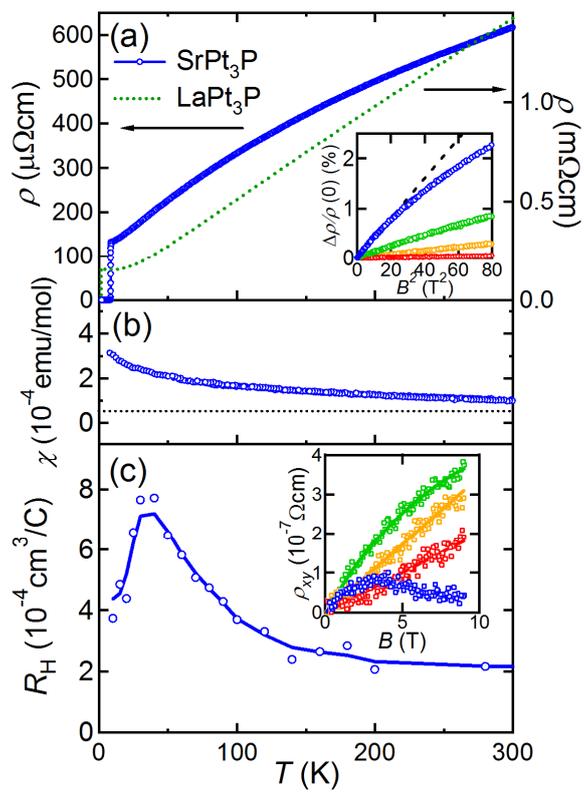

Fig. 3

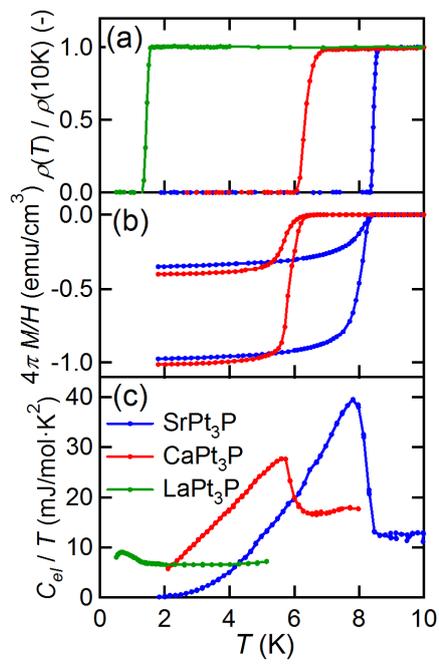

**Fig. 4**